\documentclass[aps,prd,twocolumn,preprintnumbers,dblfloatfix,nofootinbib
]{revtex4-1}
\usepackage[utf8]{inputenc}
\usepackage{amsmath}
\usepackage{graphicx}
\usepackage{amssymb}
\usepackage{color}
\usepackage{cancel}
\usepackage{framed}
\usepackage{comment}
\usepackage[margin=1in]{geometry}
\usepackage{mathtools}
\usepackage{hyperref}
\usepackage{chngcntr}

\newcommand{\Lagr}{\mathcal{L}}

\newcommand{\nn}{\nonumber}
\newcommand{\pd}{\partial}

\def\a{\alpha}  \def\g{\gamma} 
\def\d{\delta}
\def\g{\gamma} \def\m{\mu} \def\n{\nu}  \def\r{\rho}
 \def\o{\omega} \def\t{\tau} \def\l{\lambda}
\def\D{\Delta}   \def\L{\Lambda}

 \newcommand{\Ncal}{{\mathcal N}}
 \newcommand{\Ocal}{{\mathcal O}}
 
\newcommand{\Ecal}{{\mathcal E}} \newcommand{\Rcal}{{\mathcal R}}

\begin{document}

\preprint{IPMU21-0042}

\title{Probing Relativistic Axions from Transient Astrophysical Sources}

\author{Joshua Eby}
\author{Satoshi Shirai}
\author{Yevgeny V.~Stadnik}
\author{Volodymyr Takhistov}
\affiliation{Kavli Institute for the Physics and Mathematics of the Universe (WPI), The University of Tokyo Institutes for Advanced Study, The University of Tokyo, Kashiwa, Chiba 277-8583, Japan} 

\begin{abstract}
Emission of relativistic axions from transient sources, such as axion star explosions, can lead to observable signatures. 
We show that axion bursts from collapsing axion stars can be detectable over the wide range of axion masses $10^{-15} \, \textrm{eV} \lesssim m \lesssim 10^{-7} \, \textrm{eV}$ in
laboratory experiments, such as ABRACADABRA, DMRadio and SHAFT. 
The detection of axion bursts could provide new insights into the fundamental axion potential, which is challenging to probe otherwise. An ensemble of bursts in the distant past would give rise to a diffuse axion background distinct from the usual cold axion DM.
Coincidence with 
other signatures would 
provide a new window into multi-messenger astronomy. 
\end{abstract}

\date{\today}

\maketitle

%%%%%
\section{Introduction}
With the discovery of the Higgs boson, the prominent role that scalar particles play in nature has become apparent.
The quantum chromodynamics (QCD) axion~\cite{Peccei:1977hh,Weinberg:1977ma,Wilczek:1977pj,Kim:1979if,Shifman:1979if,Zhitnitsky:1980tq,Dine:1981rt} is a leading solution to the strong CP problem and also constitutes an attractive dark matter (DM) candidate~\cite{Preskill:1982cy,Abbott:1982af,Dine:1982ah}. Axions, as well as axion-like particles (ALPs) in general, are expected to be ubiquitous in string theories~(e.g., \cite{Witten:1984dg,Arvanitaki:2009fg}). 
Significant efforts have been devoted and are ongoing to explore the axion/ALP parameter space (see \cite{Irastorza:2018dyq,Agrawal:2021dbo} for a recent overview). 

Axions associated with the Galactic DM halo contributing to cold DM are understood to be non-relativistic and are the prime target of experimental efforts \cite{Irastorza:2018dyq,Agrawal:2021dbo}. 
Relativistic axions can also be readily produced through a variety of mechanisms in the early universe, potentially contributing to a cosmic axion background~(e.g., \cite{Dror:2021nyr} and references therein). 
Experimental searches for continuously-produced relativistic axions from astrophysical environments like stars have been also widely explored~\cite{Raffelt:2006cw}.

After formation in the early universe, axions could condense to form solitonic gravitationally-bound compact axion stars~(e.g., \cite{Kaup:1968zz,Ruffini:1969qy,Colpi:1986ye,Seidel:1993zk}; see e.g.~\cite{Jetzer:1991jr,Schunck:2003kk,Liebling:2012fv} for a review of boson stars in general), for example in the cores of more diffuse miniclusters \cite{Kolb:1993zz,Eggemeier:2019jsu}. 
Axion stars can naturally be associated with transient production sites of relativistic axion bursts, such as ``bosenova'' explosions~\cite{Eby:2016cnq,Levkov:2016rkk,Helfer:2016ljl}. 
Other transient astrophysical sources of axions include superradiant binary black holes (e.g., \cite{Yoshino:2012kn,Arvanitaki:2014wva,Baumann:2018vus}) and supernovae \cite{Raffelt:1999tx}. 

Here we explore the possibility of detecting relativistic axions from transient astrophysical sources using terrestrial detectors. 
We discuss both the simple case of a transient axion burst signal in the limiting case of minimal wave spreading, as well as the well-motivated case of axion star bosenovae based on a concrete production mechanism and the detailed numerical simulations of Ref.~\cite{Levkov:2016rkk}. 

We show how transient sources can lead to signatures that can be used to explore the fundamental axion potential, which is otherwise difficult to probe via conventional searches for cold axion DM. 
Further, we identify that the historic accumulation of axions from transient sources would give rise to a diffuse axion background. 
Our results are general, apply to many models and can be extended to other new physics searches from transient sources.

%%%%%
\section{Axion Star Explosions}
\label{sec:QCD}
We consider axions of mass $m$ and decay constant $f$, with the general QCD self-interaction potential \cite{diCortona:2015ldu}
\begin{equation}
\label{axion_cosine_potential}
V(\phi) = \frac{m^2 f^2(1+z)}{z}
\left[1+z-\sqrt{1+z^2+2z\cos\frac{\phi}{f}}\right] \, ,
\end{equation}
where $z\equiv m_u/m_d\approx 0.56$ \cite{Weinberg:1977hb,Zyla:2020zbs}. 
As long as the axion field $\phi\ll f$, we can focus on the leading self-interaction coupling $\l = -g_4^2\,m^2/f^2$ [where $g_4^2=(1-z+z^2)/(1+z)^2 \approx 0.3$ for the QCD chiral potential]. For the QCD axion, the scale $\L^2 =  m f \approx 6\times10^{-3}$ GeV$^2$ is fixed by the QCD confinement scale $\Lambda_\textrm{QCD} \approx 250~\textrm{MeV}$. 

At low densities, the self-interaction term is negligible and gravity can balance the gradient energy, giving rise to a stable configuration known as a ``dilute axion star'' \cite{Chavanis:2011zi,Chavanis:2011zm,Eby:2014fya,Schiappacasse:2017ham}.
The maximum mass and minimum radius allowed by stability are $M_c \approx 10 M_\textrm{P} f/(g_4m)  \approx 10^{-11} M_\odot\,f_{12}/m_5$ and $R_c \approx 0.5 g_4 M_\textrm{P}/(m f) \approx 100\,\text{ km}/(f_{12}\,m_5)$, respectively, where
we have normalized these expressions to the typical QCD axion parameter values $m_5\equiv m/(10^{-5}$\,eV$)$ and $f_{12}\equiv f/(6\times10^{11}$\,GeV$)$, with $M_\textrm{P} = 1.22\times10^{19}$\,GeV being the Planck scale. 

When the axion star mass $M$ exceeds the critical value $M_c$,
the leading attractive self-interaction destabilizes the star, triggering collapse. 
The collapse is self-similar and well-described by a non-relativistic evolution, until the final moments, when self-interactions induce relativistic decay processes in the star that deplete its mass \cite{Eby:2016cnq} as the star expands again.
The star collapses and expands $\Ocal($few$)$ more times before settling into a gravitationally-bound and non-relativistic configuration \cite{Levkov:2016rkk}. This clump may relax again into a new dilute axion star.

The differential energy spectrum $d\Ecal/dk$ of emitted axions from the collapse process has been studied via numerical simulations in \cite{Levkov:2016rkk} (see Fig.~\ref{fig:spectrum} in ~\ref{app:axionstars}).
The leading relativistic axion momentum ($k$) peak was found to be at $k/m \approx 2.4$, with less prominent peaks at higher values of $k$. 
The collapsing axion stars are found to lose $\sim 30-60\%$ of their initial mass into outgoing axion radiation, which we will show to be potentially detectable. 
The duration of the axion burst is approximately $\d t_\textrm{burst} \approx 400/m \approx 30$ ns$/m_5$ \footnote{Though not explicit in \cite{Levkov:2016rkk}, based on dimensional analysis we argue that the burst duration $\d t_\textrm{burst}$ must be approximately independent of $f$. The scaling with $m$ is given, and aside from $f$, the only other relevant dimensionful parameter present is $M_\textrm{P}$; however, gravity is practically decoupled during the final stages of collapse, as long as $f\ll M_\textrm{P}$. Therefore, $\d t_\textrm{burst}$ will be practically independent of $M_\textrm{P}$ and must therefore also be nearly independent of $f$.}. 

Further details about axion stars and the collapse simulations of \cite{Levkov:2016rkk} are reviewed in \ref{app:axionstars}.

%%%%%
\section{Axion Burst Properties}
Consider an astrophysical burst of total energy output $\Ecal$, a distance $\Rcal$ from the detector, which emits axions of fixed momentum $k_0 = q\,m$, where $q\gtrsim 1$ for relativistic axion emission. If the corresponding frequency $\omega_0$ of the outgoing axions is in the sensitive range of a given axion DM experiment, then the burst may be detectable.

As the burst travels toward the detector, wave spreading will dilute the total energy density $\rho$ as $\propto (\d t)^{-1}$, where $\d t$ is the apparent burst time as seen at the detector.
In the absence of wave spreading, this energy dilution is dictated solely by the burst duration at the source, $\d t = \d t_\textrm{burst}$, or in terms of the equivalent length scale, $\d x = \delta x_{\rm burst}$.
The burst energy density will also be diluted as $\propto \Rcal^{-2}$ due to the propagation of spherical waves away from the source. 
Altogether, in the limiting case when $\mathcal{R} \gg \delta x$, the energy density at the detector takes the form
\begin{equation} 
\label{eq:rhoE}
 \rho_* \approx \frac{\Ecal}{4\pi\,\Rcal^2 \, \d x} \, . 
\end{equation}
 
For a QCD axion bosenova of the kind simulated in Ref.~\cite{Levkov:2016rkk}, we can characterize the burst by the total energy emitted around the main relativistic momentum peak for a single explosion, which is
\begin{align}
\label{burst_peak_energy}
 \Ecal_{\rm peak} 
   &\approx 3400\,m\, \frac{f^2}{m^2} \sim 10^{41}\,{\rm GeV}\,\frac{f_{12}^2}{m_5} \, .
\end{align}
In the limit of minimal wave spreading, this implies $\rho_* \sim 10^{7}\rho_{\rm DM}f_{12}^2(100\,{\rm AU}/\Rcal)^2$, where $\rho_\textrm{DM} \approx 0.4 \, \textrm{GeV/cm}^3$ is the local DM density. 
See \ref{app:axionstars} for further details. 

The temporal properties of the burst are also affected by wave spreading effects.
In the absence of wave spreading, $\delta t = \d t_\textrm{burst}$,
whereas when wave spreading dominates, $\delta t \approx \delta k / m \times \Rcal / (q^2 \sqrt{q^2 + 1})$. 
Additionally, the apparent coherence time $\tau_*$ of the axion burst (as seen at the detector) is $\tau_* = 2\pi / \delta \omega \sim 2\pi \sqrt{q^2 + 1} / (qm)$ in the absence of wave spreading, but increases to $\tau_* \approx 2\pi \Rcal / (q^3 m \d t_\textrm{burst})$ 
in the limiting case of strong wave spreading. 
The latter follows from the consideration that, in the case of strong wave spreading, at a given time the detector is only immersed in relativistic axion waves with a small dispersion of $\delta k \approx m q^2 \sqrt{q^2 + 1} \d t_\textrm{burst} / \Rcal$. 
For further details concerning wave spreading, see ~\ref{app:wavespread}. 
Wave spreading can affect the resulting sensitivity, as we discuss below. 

In general, if objects of mass $M$ constitute a $f_{\rm DM}$ fraction of the DM, and explode on an average timescale of $\tau$, then the number of exploding objects within a distance $\Rcal$ of a detector on Earth will be $N_{\rm star}(\Rcal)=(f_{\rm DM}\r_{\rm DM}/M)(4\pi\Rcal^3/3)$, which for an axion star with mass $M=M_c$ gives $N_{\rm star}(\Rcal)\approx f_\textrm{DM}(\Rcal/100\,{\rm AU})^3(m_5/f_{12})$.
An experiment running for $t_{\rm int}=1$\,yr could thus detect at least one axion star explosion on average at a distance $\Rcal$ if 
\begin{equation} \label{eq:Ncondition}
    \Ncal \equiv N_{\rm star}(\Rcal)\times\left(\frac{\rm 1 \, yr}{\t}\right) > 1 \, . 
\end{equation}
In what follows, we set $\tau=10$\,Gyr (comparable to the present age of our Galaxy), and we ensure that our choice of $f_{\rm DM}$ satisfies gravitational lensing bounds in the mass range $10^{-11} \, M_{\odot} \lesssim M \lesssim 10 \, M_{\odot}$ \cite{Croon:2020ouk}. 
See ~\ref{app:frequency} for further discussion of axion star explosions, including the distribution of masses and frequency of explosions.

%%%%%
\section{Axion Burst Signals}
\label{Sec:axion_signals}
The transient signal from a relativistic axion burst differs from the usual DM signal associated with a cold oscillating axion field in several important aspects. 
DM axions are expected to oscillate coherently on the timescale $\tau_{\rm DM} \approx 2 \pi (m  v_{\rm DM}^2)^{-1} \sim 2 \pi \times 10^{6}/m$, where the typical virial velocity within the Milky Way DM halo is $v_{\rm DM} \sim 10^{-3}$, implying an oscillator quality factor of $Q_\textrm{DM} \sim 10^{6}$. 
On the other hand, a relativistic axion burst is composed of incoherent waves and will only have a quality factor of $Q_* = \Ocal(1)$ when the effects of wave spreading are negligible. 
As a result, the signal-to-noise ratio (SNR) in our model will generally be suppressed by some power of the ratio $Q_*/Q_\textrm{DM} \ll 1$. 
However, the burst energy density reaching a detector, $\rho_*$, can greatly exceed the local DM density. 
In contrast, for a relativistic cosmic axion background from the early universe, like that considered in Ref.~\cite{Dror:2021nyr}, the energy density is typically well below that of the cold DM component. 

The sensitivity to temporally coherent oscillating signals improves with the integration time $t_\textrm{int}$ in an experiment as $\propto t_{\rm int}^{1/2}$, up to a maximum number of oscillations dictated by the coherence time. 
In the temporally incoherent regime, the SNR still grows with the integration time, albeit more slowly, as $\propto t_{\rm int}^{1/4}$ \cite{Allen:1999stochastic,Budker:2013hfa,Grote:2019gw-dm}. 
We focus on signals that are linear in both the axion field and the coupling constant $g$. 
Since the quality factor for the axion burst signal is expected to be $Q_* = \Ocal(1)$, there is no expected benefit from resonant-type experiments, such as those described in \cite{Sikivie:1983ip}, and so we focus solely on broadband-type experiments. 

Traditional searches for axion DM usually exploit the axion's possible coupling to the electromagnetic field: 
\begin{align}
\label{Lagrangian_axion-photon_coupling}
\Lagr_\textrm{EM} =  g_{\phi\g} \, \phi \, F_{\m\n}\tilde{F}^{\m\n} \, , 
\end{align}
where $F$ is the electromagnetic field tensor and $\tilde{F}$ is its dual. 
The coupling in Eq.~(\ref{Lagrangian_axion-photon_coupling}) is derivative in nature and gives rise to an axion-induced effective electric current $\boldsymbol{j} \propto g_{\phi \gamma} (\boldsymbol{E} \times \boldsymbol{\nabla} \phi - \boldsymbol{B} \, \partial_t \phi)$, where $\boldsymbol{E}$ and $\boldsymbol{B}$ denote applied electric and magnetic fields, respectively. 
The time derivative and spatial gradient associated with a spinless field $\phi$ have the typical sizes $|\partial_t \phi| \sim \varepsilon \phi_0$ and $|\boldsymbol{\nabla} \phi| \sim k \phi_0$, respectively, where $\varepsilon = \gamma m$ is the typical particle energy ($\gamma = 1/\sqrt{1-v^2}$ is the Lorentz factor, with $v$ being the particle speed), $k = \gamma m v$ is the typical particle momentum, and $\phi_0$ is the typical field amplitude. 
The energy density associated with the field is given by $\rho \sim \varepsilon^2 \phi_0^2$. 
Broadband laboratory searches for axion DM via the electromagnetic coupling in Eq.~\eqref{Lagrangian_axion-photon_coupling} mainly search for an axion-induced oscillating magnetic flux in the presence of an applied static magnetic field \cite{Cabrera:2010-axionLC,Sikivie:2014-axionLC,Kahn:2016ABRA,Ouellet:2019ABRA,Gramolin:2021ferromagnet,Salemi:2021gck,Ouellet:2021-DMRadio}.
In this case, the sensitivity to a relativistic axion burst (with the axion-photon coupling of Eq.~\eqref{Lagrangian_axion-photon_coupling} denoted $g_*$), relative to the standard cold DM search paradigm (denoted $g_{\rm DM}$), is given by: 
\begin{equation}
\label{SNR_time-derivative_no-wave-spreading}
\frac{g_*}{g_\textrm{DM}} \sim \sqrt{\frac{\rho_\textrm{DM}}{\rho_*}} \frac{t_
\textrm{int}^{1/4} \textrm{min} \left( \tau_\textrm{DM}^{1/4} , t_\textrm{int}^{1/4} \right)}{ \min \left[ \left( \delta t \right)^{1/4} , t_\textrm{int}^{1/4} \right] \min \left( \tau_*^{1/4} , t_\textrm{int}^{1/4} \right) }  \, , 
\end{equation}
where we have made use of the inequality $\tau_* < \delta t$, with $\tau_*$ being the coherence time of the axion burst as seen at the detector. 
In deriving Eq.~(\ref{SNR_time-derivative_no-wave-spreading}), we have assumed that the apparatus is capable of sampling data points at a rate of at least $\mathcal{O}(m)$, which allows one to optimally search for transient signals in the collected data.

Axions can also couple to fermions via derivative-type couplings. 
In this case, Eq.~(\ref{SNR_time-derivative_no-wave-spreading}) is modified by the presence of an extra factor of $v_\textrm{DM} / v_* \approx v_\textrm{DM}$, which implies an enhanced sensitivity to a relativistic axion burst by the factor of $v_* / v_\textrm{DM} \sim 10^3$ compared to Eq.~(\ref{SNR_time-derivative_no-wave-spreading}). 
However, we are not aware of sufficiently sensitive broadband techniques for the relevant mass range in this case. 
We discuss the axion-fermion coupling and some other types of couplings in more detail in \ref{app:SD} and \ref{app:ND}. 

As a consequence of the relations above, if $\d t<t_\textrm{int}$, then the SNR remains the same regardless of the degree of wave spreading, since the signal can be captured in its entirety. 
On the other hand, if $\delta t > t_\textrm{int}$, then the experimental sensitivity will degrade in accordance with Eq.~(\ref{SNR_time-derivative_no-wave-spreading}). 
We also remark that the comparison in Eq.~\eqref{SNR_time-derivative_no-wave-spreading}
involves a fixed value of $\o_0$, rather than  $m$. 
Since cold DM axions have $\o_0 \approx m$, whereas axions in bursts are relativistic with $\o_0 \gg m$, experiments searching for relativistic axion bursts can therefore be sensitive to lower axion masses compared to cold DM searches at the same signal frequency.

\begin{figure}[t]
 \centering
 \includegraphics[scale=0.5]{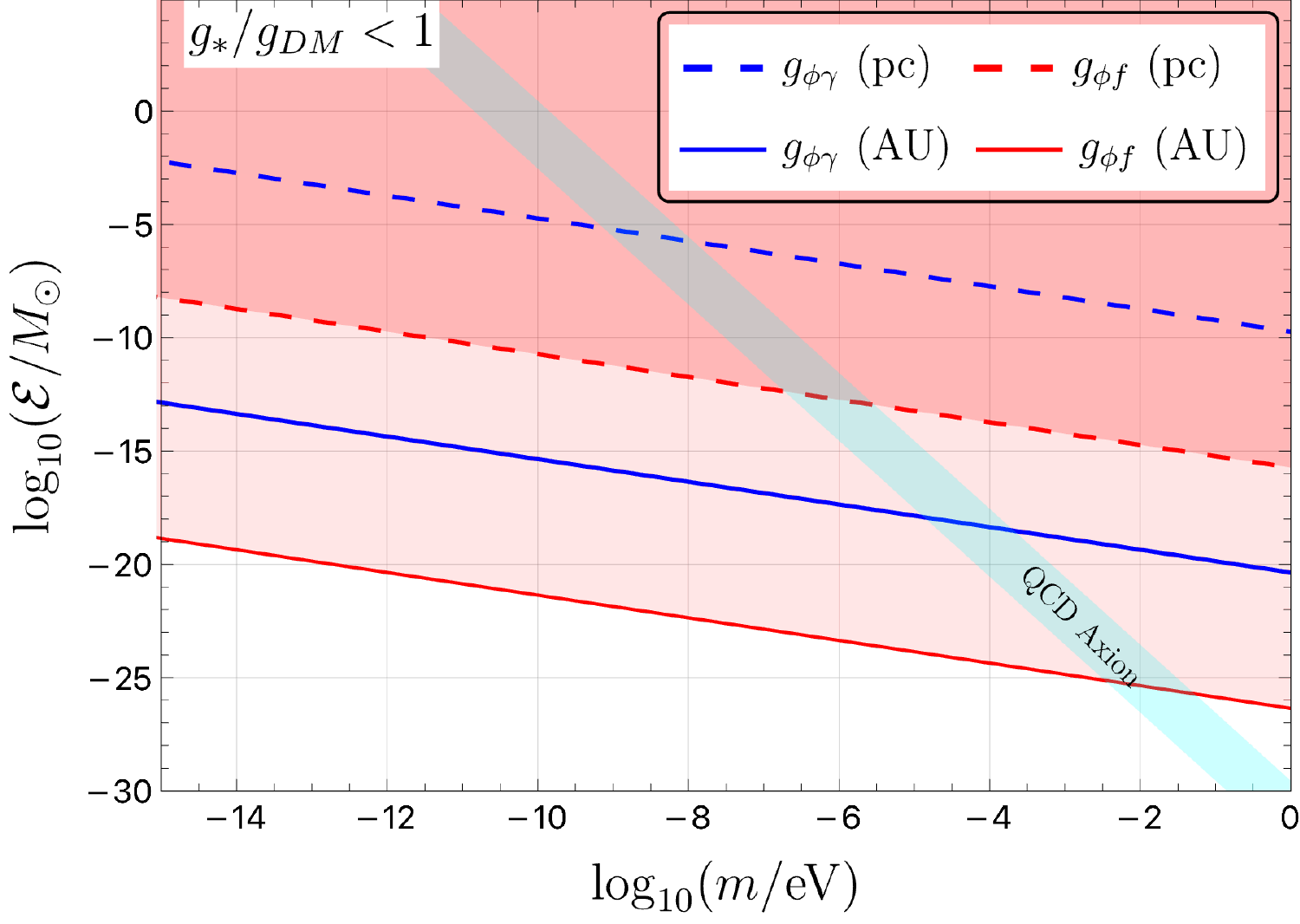}
 \caption{
Regions of parameter space for a minimum-uncertainty burst [$\d k \approx (\d x_{\rm burst})^{-1}$] for which $g_*/g_\textrm{DM}<1$ for $g_{\phi\gamma}$ [blue lines, using Eq.~\eqref{SNR_time-derivative_no-wave-spreading}] and for $g_{\phi f}$ [red lines, using Eq.~\eqref{SNR_spatial-derivative_no-wave-spreading}].
 The shaded regions indicate where bursts of total energy $\Ecal$ at the distances $\Rcal=$ 1\,AU and 1\,pc (solid and dashed lines, respectively) would give rise to greater signal strength in a detector for a relativistic axion burst as compared with conventional experiments searching for cold axion DM, provided that at least one axion bosenova occurs during the course of the experimental integration time of $t_{\rm int} =$ 1\,yr.
 The lines assume $k_0/m=100$ (for $\Rcal\lesssim$ kpc, the dependence on $k>1$ is not significant).
}
 \label{fig:ToySens}
\end{figure}

\begin{figure*}[t]
 \centering
  \includegraphics[width=0.47\textwidth]{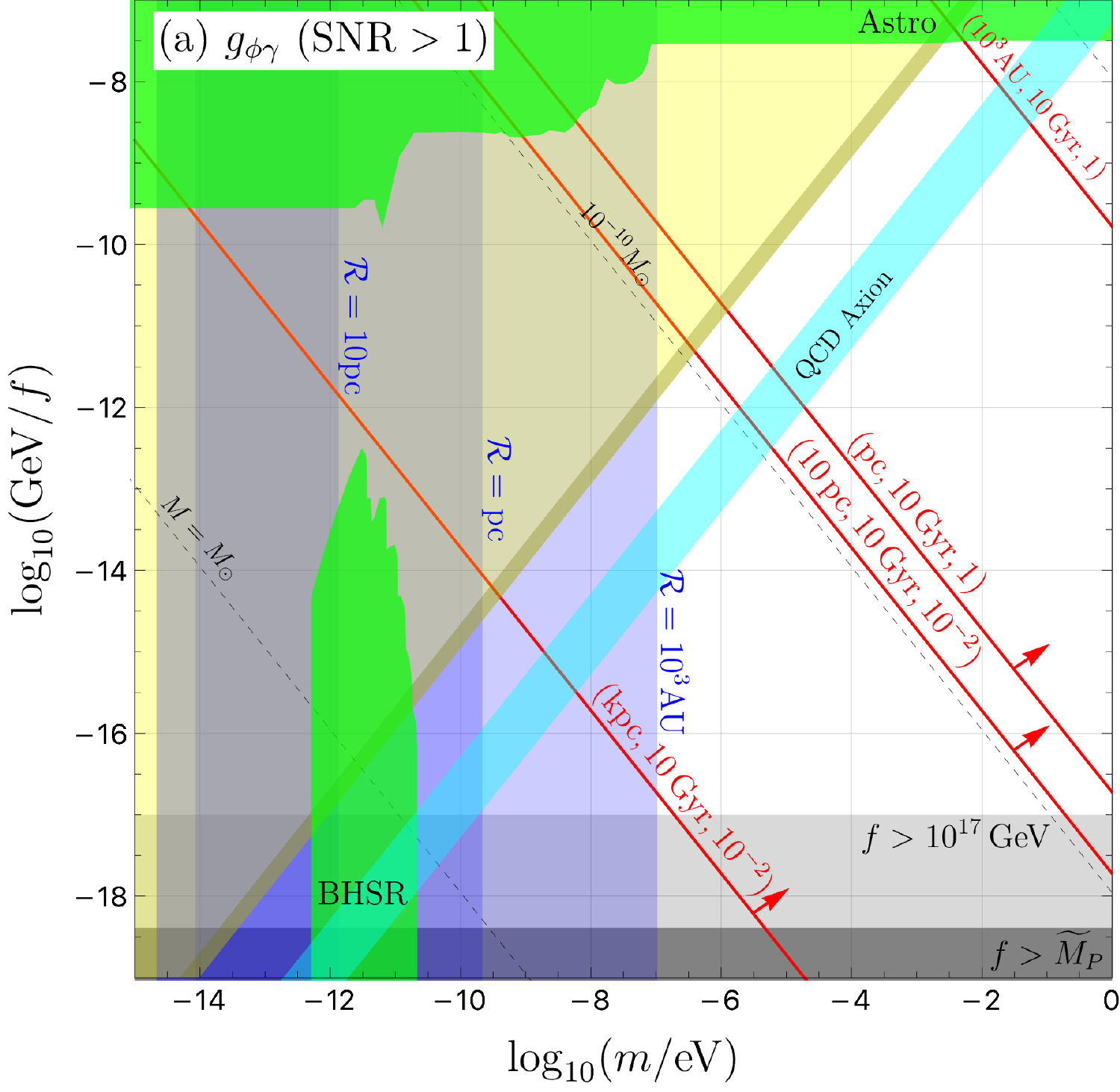}\quad
  \includegraphics[width=0.47\textwidth]{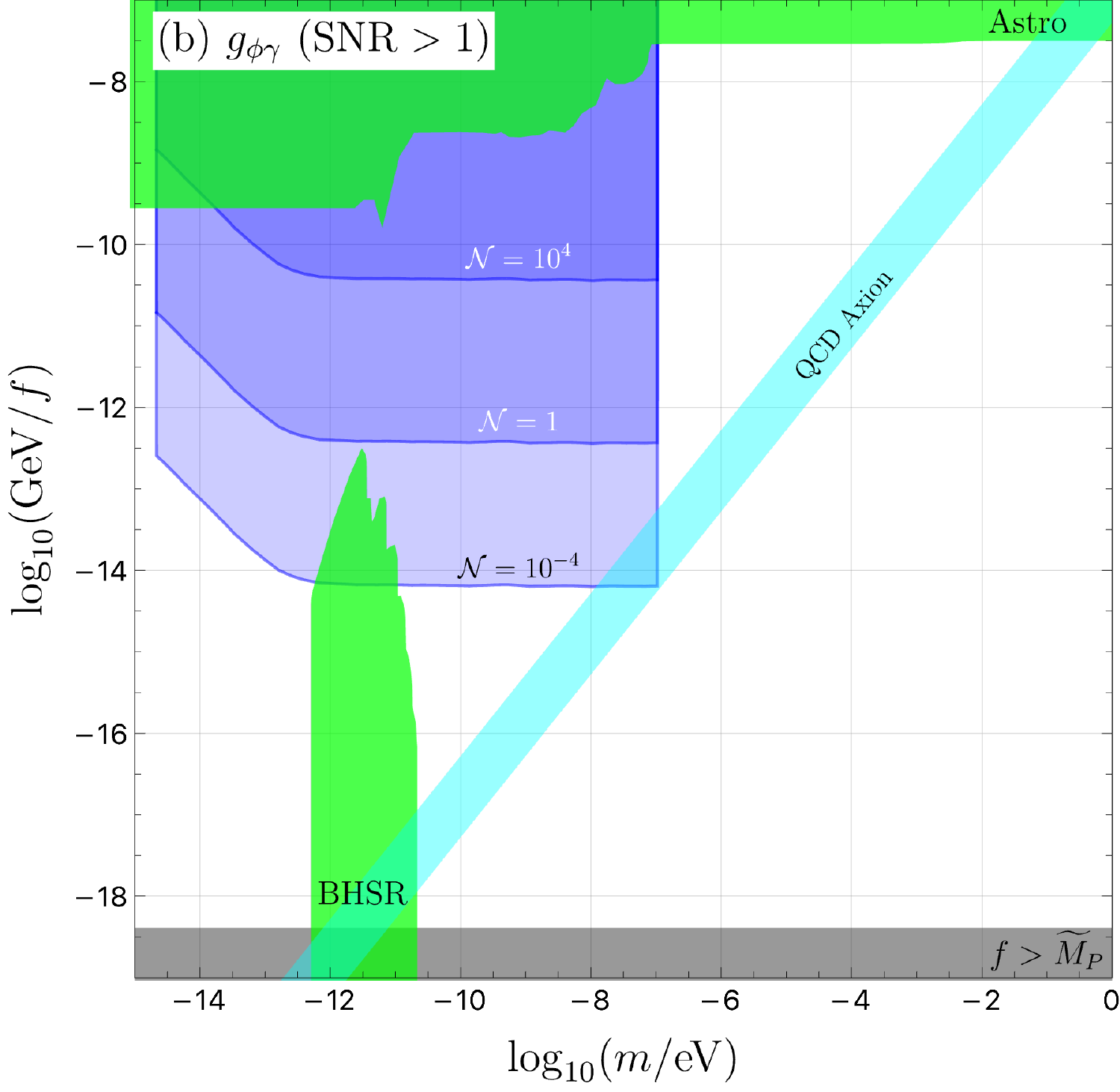}
\caption{
  \label{fig:QCDsens}
\textbf{(a)} Sensitivity reach of an ABRACADABRA-type detector (in its $100$\,m$^3$ final projected form \cite{Kahn:2016ABRA}) to axion star bosenovae using Eq.~\eqref{SNR_time-derivative_no-wave-spreading} (blue regions), for the different explosion distances $\Rcal=10^3\,\mathrm{AU}, 1\,\mathrm{pc}, 10\,\mathrm{pc}$. 
 The QCD axion band is illustrated in cyan. 
 The red lines mark $\Ncal \ge 1$ for different choices of the three relevant parameters $\{\Rcal,\tau,f_\textrm{DM}\}$ as labelled, chosen to satisfy gravitational lensing bounds \cite{Croon:2020ouk}, and the black dashed contours mark fixed values of the axion star mass $M$. 
 The green regions are excluded by astrophysical observations \cite{DiLuzio:2021gos} or by the non-observation of black hole superradiance (BHSR) in rapidly-rotating black holes \cite{Arvanitaki:2014wva}. For concreteness, we have use the model-dependent expression $|g_{\phi\g}| = 1.92\,\a / (2\pi f)$ to translate coupling strength in the experimental limits to the vertical axis scale $1/f$.
 The gray regions denote $f>10^{17}$ GeV (light gray) from the consideration of axion star instability  \cite{Eby:2020ply}, and $f>\widetilde{M}_\textrm{P} = 2.4\times10^{18}$ GeV (dark gray) from theoretical considerations \cite{Montero:2015ofa} (though see also \cite{Kaplan:2015fuy,Fonseca:2019aux}).
 Finally, the yellow regions indicate where parametric resonance conversion of axions to photons would be relevant, either during axion star collapse (dark yellow) or for stable axion stars at $M=M_c$ (light yellow) \cite{Levkov:2020txo}. 
 \textbf{(b)} Sensitivity reach of an ABRACADABRA-type detector to generic ALP star explosions. 
 The blue lines are obtained by varying $\Ecal$ and $\Rcal$ to obtain the strongest signal at fixed $\Ncal$: $\Rcal=$~1\,pc and $\Ecal=4\times10^{-16}M_\odot$ with $f_\textrm{DM}=1$ ($\Ncal=10^{4}$, upper line); $\Rcal=$~1\,pc and $\Ecal=4\times10^{-12}M_\odot$ with $f_\textrm{DM}=1$ ($\Ncal=1$, middle line); and $\Rcal=$~1\,kpc and $\Ecal=4M_\odot$ with $f_\textrm{DM}=10^{-1}$ ($\Ncal=10^{-4}$, lower line).
  }
\end{figure*} 

In Fig.~\ref{fig:ToySens}, we estimate the regions of parameter space where the sensitivity ratio $g_{\ast}/g_{\rm DM} < 1$ using Eq.~\eqref{SNR_time-derivative_no-wave-spreading} for the photon coupling, and including a $v_* / v_{\rm DM}\sim 10^3$ enhancement factor for the fermion coupling (see ~\ref{app:SD}).
We focus on the optimal case of a minimal-uncertainty burst [$\d k \approx (\d x_{\rm burst})^{-1}$] of very short duration ($\d t_{\rm burst} = 2\pi/m$), and we assume $t_{\rm int} = 1$ yr.
Since this is a ratio of sensitivities, the result is independent of the details of a particular broadband-type experiment. 
It is intriguing that, for both types of couplings, the sensitivity ratio favors the detection of relativistic axion bursts over a wide swath of feasible parameter space (e.g., for $\Ecal \lesssim M_\odot$ and $\Rcal \gtrsim $ pc). 

For the specific case of the QCD axion star collapse \cite{Levkov:2016rkk},
we use $\Ecal$ as given by Eq.~\eqref{burst_peak_energy} and depict the result by the light blue shaded region in Fig.~\ref{fig:ToySens}.
[Note that in practice, for the axion star bursts we consider here, $\d t \lesssim 1$ yr for the main relativistic peak at $q \approx 2.4$, as long as $\Rcal \lesssim \Ocal(10) \, \textrm{pc}$.] 
When the ratio $g_*/g_\textrm{DM} < 1$ at some value of $\o_0$, this implies the possibility of the detection of an axion star bosenova with enhanced sensitivity over analogous axion cold DM searches, provided that at least one axion bosenova occurs during the experimental measurements. In fact, we will see that QCD axion stars on the KSVZ line can potentially be probed by searching for relativistic axion bursts.

In Fig.~\ref{fig:QCDsens}(a), 
we illustrate the axion-photon coupling $g_{\phi \gamma}$ sensitivity reach of a future ABRACADABRA-type \footnote{ABRACADABRA is in the process of merging with the DMRadio collaboration \cite{Salemi:2021gck,Ouellet:2021-DMRadio}; for simplicity, we use the long-term projection in \cite{Kahn:2016ABRA} to estimate the sensitivity to our signal of interest. For the current ABRACADABRA limits, see \cite{Ouellet:2019ABRA,Salemi:2021gck}; see also the related limits from SHAFT~\cite{Gramolin:2021ferromagnet}.} instrument to our axion star bosenova signal (using Eq.~\eqref{burst_peak_energy} for the emitted energy as before), along with existing constraints, for the source distances $\Rcal=10^3$\,AU, 1\,pc, $10$\,pc. 
The translation of the parameter $g_{\phi\g}$ to $1/f$ in Fig.~\ref{fig:QCDsens}(a) is model dependent; for concreteness, we have used $|g_{\phi\g}| = 1.92\,\a / (2\pi f)$, which is consistent with the KSVZ axion model when the ratio of the electromagnetic and colour anomaly coefficients is given by $E/N = 0$. The sensitivity of conventional cold axion DM searches falls off as $1/f$ (or faster), because the axion-to-standard-model couplings are proportional to $1/f$. 
On the other hand, the energy emitted in axion star bosenovae generally \emph{grows} in size as $\propto f$, since $\Ecal \propto M_c \propto f$;
therefore, the product of the two factors is \textit{independent} of $f$ for axion star burst signals in ABRACADABRA-type detectors, and hence the sensitivity regions in Fig.~\ref{fig:QCDsens}(a) are vertical lines in the $f - m$ plane. 
Axion star bosenovae may thus be a preferred method of discovery for large values of the decay constant $f$.

In Fig.~\ref{fig:QCDsens}(b),
we consider the sensitivity of a future ABRACADABRA-type instrument to explosions of arbitrary $\Ecal$, taken equal to the DM object mass $M$. 
Fixing the expected number of axion star explosions in 1 year, $\Ncal$, implies a relationship between $\Rcal$ and $M=\Ecal$ at fixed $\tau$ and $f_\textrm{DM}$; 
we vary $\Ecal$ and show the most optimistic reach of an ABRACADABRA-type detector. 
In the estimation, we assume a minimum-uncertainty and short-duration burst, as in Fig.~\ref{fig:ToySens}. 
We observe that generic axion star bursts can be both detectable (SNR $>1$) and occur frequently ($\Ncal > 1$) over a wide parameter space. We include for reference a region for $\Ncal = 10^{-4}$, which could be relevant for a scenario in which $\tau$ is much longer than our assumption of $10$ Gyr.

%%%%%
\section{Implications}
In models of cold oscillating axion DM with the cosine potential \eqref{axion_cosine_potential}, the leading-order quadratic term of the potential gives rise to a cosinusoidal time-varying function of amplitude $\phi_\textrm{LO} = \phi_0$ and angular frequency set by $m$. 
Higher-order terms in the cosine potential (\ref{axion_cosine_potential}) modify the purely cosinusoidal-in-time function into a Jacobi elliptical function due to the effects of anharmonicity.
The next-to-leading-order quartic term in Eq.~(\ref{axion_cosine_potential}) induces a correction to the leading-order cosinusoidal time-varying solution of the order of $\delta \phi_\textrm{NLO} / \phi_\textrm{LO} \sim \phi_0^2 / f^2 \sim \rho / (m f)^2 \sim 10^{-37}/(m_5 f_{12})^2$, 
assuming that the oscillating axion field saturates the local cold DM density $\rho_\textrm{DM}$. 
Unlike traditional cold DM searches, the signal from axion star bursts can only arise in the presence of axion self-interactions of at least next-to-leading order ($\phi^4$) in the axion potential, such as in Eq.~(\ref{axion_cosine_potential}). 
Furthermore, in the final stages of axion star collapse, the underlying processes are relativistic, and so one would expect next-to-next-to-leading-order ($\phi^6$) and higher-order terms in the axion potential to have a non-negligible effect on the structure of the axion emission spectrum in Fig.~\ref{fig:spectrum}. Therefore, the detection of such bursts could provide insight into the fundamental axion potential, which is challenging to probe via conventional cold DM searches. 

In analogy with the diffuse neutrino background (e.g., \cite{Beacom:2010kk,Horiuchi:2008jz,Munoz:2021sad}), relativistic axions from historic transients will accumulate into a diffuse axion background. 
The transient event rate would depend on the considered source formation and axion emission model. Highly redshifted axions could become non-relativistic, while the boosted diffuse axion background component is expected to have a distinct phase space distribution.

Depending on the specific axion coupling, a variety of multi-messenger signatures accompanying relativistic axion bursts could be potentially expected from transient sources. 
This includes radio-photon emission via the axion-photon coupling (e.g., \cite{Tkachev:2014dpa}), see the yellow regions in Fig.~\ref{fig:QCDsens}(a), as well as gravitational waves in the case of binary mergers or asymmetric axion star explosions. 
Such coincidence signatures would provide a complementary handle for exploring relativistic axions from transients. 
Searching for correlated signatures using a network of spatially-separated detectors would allow for the localization of the source. 
Some aspects of the detection prospects of relativistic axions and the multi-messenger aspect have been considered recently in the context of neutron star and black hole mergers~\cite{Dailey:2020sxa}, as well as neutron star -- axion star collisions~\cite{Dietrich:2018jov}. 
Further, it is expected that a typical axion star explodes several times before finally settling in a dilute, gravitationally-bound configuration~\cite{Levkov:2016rkk}.
Additionally, relativistic axions may convert into photons in Earth's ionosphere (see, e.g., \cite{Esteban:2019hcm}).
We leave the detailed exploration of such signatures for future work. 

Finally, we have pointed out that resonant-type experiments are less advantageous for axion burst signals than are broadband searches. 
However, this can be partly ameliorated by the use of simultaneous resonant searches in a narrow range of frequencies, which allows for the capture of a greater fraction of the burst energy and for probing a larger region of the emission spectrum. 
This constitutes a middle-ground approach between purely resonant and purely broadband search strategies. 
More generally, the broadness of the axion burst signal provides an important way to distinguish it from a cold DM signal (which would be very narrow in frequency space). 
A full treatment of the signal shape for axion bosenova signals, including the effect of multiple subsequent explosions as found in \cite{Levkov:2016rkk}, is left for future work.

%%%%%
\section*{Acknowledgments}
We thank D.~G.~Levkov for helpful discussions. 
This work was supported by the World Premier International Research Center Initiative (WPI), MEXT, Japan and by the JSPS KAKENHI Grant Numbers JP20K14460 (Y.V.S.) and 17H02878, 18K13535, 20H01895, 20H05860 and 21H00067 (S.S.). 

%%%%%%%%%%%%%%%%%%%%%%%%%%%%%%%%%%%%%%%%%

\appendix

%%%%%
\section{Basic Properties of Axion Stars} 
\label{app:axionstars}

Axion stars are characterized by a classical field wavefunction in the non-relativistic limit.
For a Klein-Gordon field $\phi$, we define \cite{Guth:2014hsa}
\begin{equation}
 \phi(r,t) = \frac{1}{\sqrt{2m}}\left[e^{-i\,m\,t}\psi(r,t) + {\rm h.c.}\right] \, , 
\end{equation}
and the equation of motion for $\psi(r,t)$ is a nonlinear Schr\"odinger equation 
\begin{equation}
 i\,\dot\psi =\left[-\frac{\nabla^2}{2m} + V_g(|\psi|^2) - \frac{g_4^2\,|\psi|^2}{8\,f^2}\right]\psi \, , 
\end{equation}
at leading order in the self-interaction potential. The gravitational potential $V_g(|\psi|^2)$ is defined by
\begin{equation}
 \nabla^2 V_g = 4\pi\,G\,m^2\,|\psi|^2 \, ,
\end{equation}
and the total mass is defined by $M\equiv m \int d^3r\,|\psi|^2$.

At low densities, the self-interaction term is negligible and gravity can balance the gradient energy, giving rise to a stable configuration known as a ``dilute axion star" \cite{Chavanis:2011zi,Chavanis:2011zm,Eby:2014fya,Schiappacasse:2017ham}. 
Since axion star wavefunctions are non-compact, we define the effective radius as $R=R_{99}$, inside of which a mass $0.99M$ is contained. 
Solutions to the equations of motion in the dilute limit satisfy
\begin{equation} \label{eq:RM}
 R = R_c\left(\frac{M_c}{M}\right) \left[1 + \sqrt{1 - \frac{M^2}{M_c{}^2}}\right] \, ,
\end{equation}
with the maximum mass (minimum radius) allowed by structural stability\footnote{This is referred to by other authors as ``gravitational stability", although the instability observed when $M\simeq M_c$ is induced by self-interactions rather than gravity.} of the star given by 
\begin{align} \label{eq:McRc}
 M_c &\approx 10\frac{M_\textrm{P} f}{g_4\,m} 
    \approx 10^{-11}
 			M_\odot 
    \frac{f_{12}}{m_5} \, ,  \\
 R_c &\approx 0.5\frac{g_4 M_\textrm{P}}{m f}
    \approx \frac{100 \text{ km}}{f_{12}\,m_5} \, .
\end{align}

\begin{figure}[t]\hspace{-8mm}
\includegraphics[scale=0.55]{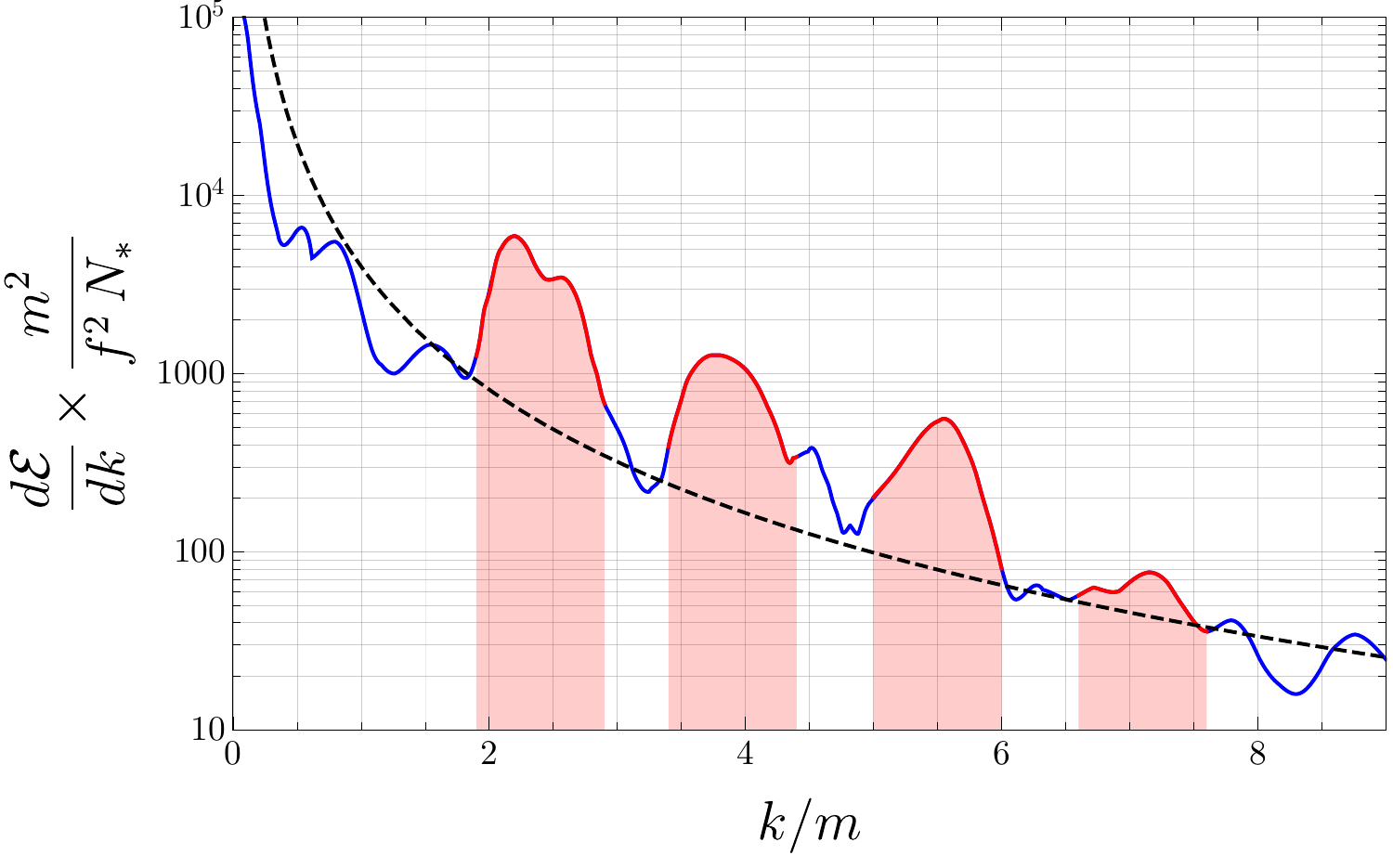}
 \caption{Spectrum of emitted axions in a Bosenova explosion from a collapsing axion star (solid line), reproduced from Fig.~3a of \cite{Levkov:2016rkk}.  
 A rough power-law fit $d\Ecal/dk \propto k^{-2.3}$ is given as a guide (dashed black line). 
 Peaks of width $\D k\approx m$ centered around $k/m \approx 2.4,3.9,5.5,7.1$ are shaded in red. 
 The parameter $N_*$ denotes the number of explosions during the collapse.
  } 
\label{fig:spectrum}
\end{figure}

When axion stars exceed their critical mass $M_c$, they collapse and explode in a bosenova of relativistic particles, a process simulated in Ref.~\cite{Levkov:2016rkk}. We reproduce the emitted axion spectrum in Fig.~\ref{fig:spectrum}. 
In \cite{Levkov:2016rkk}, it was shown that after $N_*=\Ocal($few$)$ explosions, the axion star settles into a diffuse, gravitationally-bound configuration; in the end, the total energy loss of the star is well-fit by the linear function $\Ecal_{\rm loss}/M_c \approx 0.3 + 830 \, f/M_\textrm{P}$ for $f/M_\textrm{P} \lesssim 10^{-3}$. 

To get a rough estimate of the signal from this bosenova, suppose that a collapse and bosenova of the kind simulated in Ref.~\cite{Levkov:2016rkk} occurs a distance $\Rcal$ from a terrestrial detector. 
The integral over the main relativistic peak in Fig.~\ref{fig:spectrum}, centered around $k/m \approx 2.4$, gives
\begin{equation}
\frac{\Ecal_{\rm peak}}{m} \approx \int_{1.9}^{2.9}\,d(k/m) \frac{d\Ecal}{dk} \approx 3400\,N_*\,\frac{f^2}{m^2} \, ,
\end{equation}
which we use in the calculation of  Eq.~\eqref{burst_peak_energy} in the Main Text, with $N_*=1$. Higher-momentum peaks are relevant to the signal as well, though we leave a full analysis of the signal shape for future work. Note that the largest proportion of emitted axions are found in the non-relativistic region of the spectrum, particularly at $k/m\lesssim 0.5$. It would be interesting to consider such axions as a transient, non-relativistic, DM-like signal, although this is beyond the scope of the present work, which focuses on relativistic axion bursts. Note however that very non-relativistic axions emitted from a burst would, given enough time, become indistinguishable from the usual cold DM.

The duration of the axion star burst for $f = 2 \times 10^{-4} M_\textrm{P} = 2.4 \times 10^{15}$ GeV in the simulation of \cite{Levkov:2016rkk} was approximately $\d t_\textrm{burst} \approx 400/m \approx 30$ ns/$m_5$.
Since the emitted axions are relativistic, the axion burst will have a corresponding intrinsic spread in space of $\d x_{\rm burst} \approx \d t_\textrm{burst} \approx 10\,\textrm{m} / m_5$ along the direction of propagation. 
In this approximation, substituting Eq.~\eqref{burst_peak_energy} into Eq.~\eqref{eq:rhoE}, we obtain: 
\begin{align} \label{eq:rhoplus}
 \rho_{*} &\approx \frac{\Ecal_\textrm{peak}}{4\pi\,\Rcal^2 \, \d x_{\rm burst}} 
			\sim 10^7\,\frac{\rm GeV}{{\rm cm}^3}\,f_{12}^2 \left(\frac{100\,{\rm AU}}{\Rcal}\right)^2 \, , 
\end{align}
which does not depend on the size of the terrestrial detector. 
This will be modified by wave spreading effects, see \ref{app:wavespread}. 

%%%%%
\section{Wave Spreading Effects}
\label{app:wavespread}

Our signal depends on the duration of the burst, $\d t$, as seen at the detector, which in turn depends on the spread in momentum of the emitted axion particles at the source.
A true $\d$-function power spectrum in frequency space is impossible, as the power spectrum cannot be infinitesimally narrow without violating the uncertainty principle; the spread in momentum $\d k$ will be at least of order $(\d x_\textrm{burst})^{-1}$. 
Even a small spread in the emitted axion momenta can be important, as it leads to a further dilution of the axion energy density at the position of Earth due to wave spreading (in the particle picture, the fastest axions reach Earth sooner than slower ones, even if they are emitted at the same time). 
This is particularly important if an explosion occurs sufficiently far away. Here we derive the relevant time/distance scales needed for the analysis in the Main Text.

Suppose the axion momentum has a spread $\d k$ around some central value $k_0$. 
The relativistic energy-momentum relation is $\omega = \sqrt{k^2 + m^2} = m \sqrt{q^2+1}$, with $q=k/m$, which implies $v^2 = q^2/(q^2 + 1)$. 
Taking the differentials of both sides, we obtain
\begin{align} \label{eq:dv}
 &v\,\d v = \frac{q\,\d q}{(1+q^2)^2} \nn \\
 & \qquad \Rightarrow \qquad \frac{\d v}{v} 
 					= \frac{\d q}{q}\frac{1}{1+q^2} = \frac{\d k}{k} \frac{m^2}{\omega^2} \, . 
\end{align}
Eq.~\eqref{eq:dv} is relevant because the wave spreading effect is proportional to $\d v/v$. 
In particular, for an instantaneous burst with small momentum spread $\d k/k \ll 1$ around the central peak at $k_0$, we find
\begin{equation} \label{eq:dx1}
 \d x = \frac{\d v}{v}\,\Rcal 
   \approx \frac{\d k}{k_0} \frac{m^2}{k_0^2 + m^2} \Rcal \, .
\end{equation}
This implies
\begin{equation}
    \delta t \approx \frac{\delta k}{m} \frac{\Rcal}{q^2 \sqrt{q^2 + 1}} \, , 
\end{equation}
using $\d t = \d x \sqrt{q^2+1}/q$; this expression is used in the results of the Main Text in Eq.~\eqref{SNR_time-derivative_no-wave-spreading}.

The smallest momentum spread at the source that is allowed by the uncertainty principle is $\d k \approx (\d x_{\rm burst})^{-1}$, which implies
\begin{equation} \label{eq:dx2}
    \frac{\d x}{\d x_\textrm{burst}} \gtrsim \frac{\Rcal}{ q \left( q^2+1 \right) m \left(\delta x_\textrm{burst}\right)^2 } \, . 
\end{equation}
So the intrinsic burst duration is negligible as long as 
$\Rcal/\d x_\textrm{burst} \gg q(q^2+1)\,m\,\d x_\textrm{burst}$, which holds unless $q$ is very large or $\Rcal$ is very small. 

Let us now consider the special case of axion star bosenovae. 
As we have just demonstrated, wave spreading in flight can often dominate the intrinsic signal duration. 
When the burst duration is set by the axion mass scale $m$, we can write $\d x_\textrm{burst} = \xi/m$, where $\xi$ is a dimensionless constant in natural units, in which case Eq.~\eqref{eq:dx2} becomes:
\begin{equation}
    \frac{\d x}{\d x_\textrm{burst}} \gtrsim 
    \frac{1.6\times10^{13}}{\xi^2\,q(q^2+1)}
    \left(\frac{\Rcal}{{\rm pc}}\right)
    \left(\frac{m}{10^{-10}\,{\rm eV}}\right) \, . 
\end{equation}
For the specific case of a bosenova explosion of a QCD axion star, for which $\xi \approx 400$, with $\delta k / m \sim 1$ and $q \approx 2.4$ for the main relativistic peak in Fig.~\ref{fig:spectrum}, the wave spreading effect dominates over the intrinsic spread for any burst outside of a small radius $\Rcal \gtrsim 0.03$ pc when $m \gtrsim 10^{-15}$ eV.

%%%%%
\section{Axion Star Collapse Frequency}
\label{app:frequency}

A thorough treatment of the distribution of axion stars and collapse rate is beyond the scope of this work. 
Here, we have parametrised the relevant effects via three constants: 
(a) the fraction $f_\textrm{DM}$ of DM axions contained in axion stars at the present day; 
(b) the peak value $f_c = M/M_c$ of the (unknown) mass distribution of stable axion stars in our Galaxy; 
and (c) the typical timescale $\t$ for an axion star to grow in mass to the critical point and collapse. 
In lieu of a detailed study of axion cosmology, we comment on each of these points below:

$\mathbf{1.}$ Cosmological simulations of the axion field have recently been performed by several independent groups \cite{Schive:2014dra,Mocz:2017wlg,Eggemeier:2019jsu,Nori:2020jzx}. 
    For the QCD axion, it has been shown that if the axion global symmetry (e.g., the Peccei-Quinn symmetry \cite{Peccei:1977hh}) is broken after inflation, overdensities known as axion miniclusters naturally form from field fluctuations shortly before matter-radiation equality. 
    Quantitative results, including the spectrum of overdensities, depend on very sensitive details of global string and domain wall networks in the early universe, a topic of heated debate of late \cite{Yamaguchi:1998gx,Gorghetto:2018myk,Vaquero:2018tib,Buschmann:2019icd,Gorghetto:2020qws}.
    Still, a qualitative picture is emerging in which these miniclusters form in the early universe and subsequently undergo mergers and possibly tidal destruction inside of galactic DM halos, and those that survive seed the formation of axion stars \cite{Kolb:1993zz,Levkov:2018kau,Eggemeier:2019jsu,Kirkpatrick:2020fwd}.
    
    In direct simulations of axion minicluster halos, it has been shown that a large fraction of order $\sim 0.75$ of the axion DM density remains in miniclusters down to redshift $z \approx 100$ \cite{Eggemeier:2019khm}. At lower redshift, uncertainties about merger histories and tidal stripping make it very difficult to determine the present-day distribution of these miniclusters (for recent efforts, see \cite{Kavanagh:2020gcy,Xiao:2021nkb}). 
    There are also so-called hybrid DM scenarios in which axions and primordial black holes both constitute a sizeable fraction of DM; in that case, axions can form axion stars in the gravitational potential of black holes directly from the galactic background \cite{Hertzberg:2020hsz}. 
    On the basis of the above, it is not implausible to expect axion stars to constitute a few percent or more of the total DM abundance in galaxies.

To estimate the maximum possible reach, we would like to consider the case when $f_\textrm{DM} = 1$. 
    On the other hand, gravitational lensing experiments rule out axion stars with large values of $f_\textrm{DM}$ in some range of axion star masses. 
    In the mass range $10^{-11} \, M_\odot \lesssim M \lesssim 10 \, M_\odot$, the constraint is as strong as $f_\textrm{DM} < 5\times 10^{-3}$, but can also be significantly weaker over that mass range \cite{Croon:2020ouk}. 
    In our rate estimates, we check to ensure that the abundance assumed is consistent with the gravitational lensing bounds at each value of $M$ that we consider.
    
$\mathbf{2.}$ The initial mass distribution of axion stars depends on the mechanism of their formation. 
    For example, axion stars forming inside of miniclusters have a mass at formation determined by the density and virial velocity of their host minicluster \cite{Eggemeier:2019jsu}. 
    Furthermore, once axion stars form (either from relaxation of miniclusters or some other mechanism), they can merge \cite{Palenzuela:2006wp,Cotner:2016aaq,Bezares:2017mzk,Eggemeier:2019jsu} or accrete further mass from the axion background \cite{Chen:2020cef}. 
    Therefore, the mass distribution of axion stars will be ever-changing.
    
    However, axion star masses do not grow without bound; as we have described, when the mass reaches $M_c$, the axion star collapses and emits a large fraction $\Ecal/M_c \sim 0.3-0.6$ of its mass as relativistic particles \cite{Levkov:2016rkk}. 
    Therefore, it is plausible that axion star masses will be clustered in some narrow range within an order of magnitude of $M_c$. 
    In this work, we make the simplifying assumption that all axion stars have $f_c \equiv M/M_c = 1$, in order to estimate the number of nearby axion stars and their mass.
    
$\mathbf{3.}$ In order for axion stars to be seen exploding in the sky near enough to Earth, it is necessary to ask how long it would take a typical axion star to grow enough mass (through accretion and mergers) to trigger collapse. 
    The situation is actually more complex than mere mass growth: after an initial bosenova, simulations suggests the non-relativistic remnant of the original axion star remains gravitationally bound, though it retains only $\sim 40-70\%$ of its original mass and the axions in the resulting diffuse configuration are not in their ground state \cite{Levkov:2016rkk}. 
    We speculate, therefore, that the final axion configuration may relax again into a sub-critical axion star; if this is true, then this new star will begin to accrete mass again, growing towards $M=M_c$, until it collapses and explodes again. 
    These processes might occur repeatedly in galactic haloes, leading to a higher frequency of axion star bosenovae. 
    
    All of these complex dynamical processes deserve a dedicated treatment, but for the purposes of the present paper, we parametrise the overall result by a single timescale $\tau$, the axion star lifetime, defined as the average total time for an average star to accrete mass, explode, reform into a gravitationally-bound relic, and relax again into a stable star. 
    If $\tau$ is much greater than the lifetime of the Universe, then axion stars would be effectively stable and bosenovae will be exceedingly rare.
    On the other hand, if $\tau$ is too small, then DM haloes could become unstable; every explosion would convert an $\Ocal(1)$ fraction of the axion star's mass-energy content (which may be associated with the cold DM) to relativistic axions that escape the galaxy \footnote{This may imply constraints on axion DM forming axion stars directly from the observed stability of our Galactic DM halo on cosmologically long timescales, a topic we will return to in a future work.}. 
    Therefore, in the present paper, we consider the interesting intermediate value $\tau = 10$ Gyr, which is comparable to the present age of our Galaxy.  
    
    It is crucial to emphasize that $\tau$ is highly model-dependent, and is worthy of a more complete treatment than we have offered here. For axion stars forming within axion miniclusters, for example, it has been suggested recently \cite{Eggemeier:2019jsu,Chen:2020cef} that mass growth of an axion star by accretion saturates at some mass $M_{\rm sat} < M_c$, constraining this possible mechanism for collapse at $M\simeq M_c$ (see also Section V of \cite{Yavetz:2021pbc} for the case of ultralight axions). However, this too is somewhat model-dependent; for QCD axions with $f\lesssim 10^{11}$ GeV (or ALPs with comparable self-interaction strength), the stars saturated in denser configurations, effectively driving $M\to M_c$ prior to saturation and triggering collapse as we suggest. We leave further discussion of these issues for future work.

%%%%%
\section{Derivative Coupling of Axion to Fermions}
\label{app:SD}

\begin{figure}[t]
 \centering
 \includegraphics[scale=0.5]{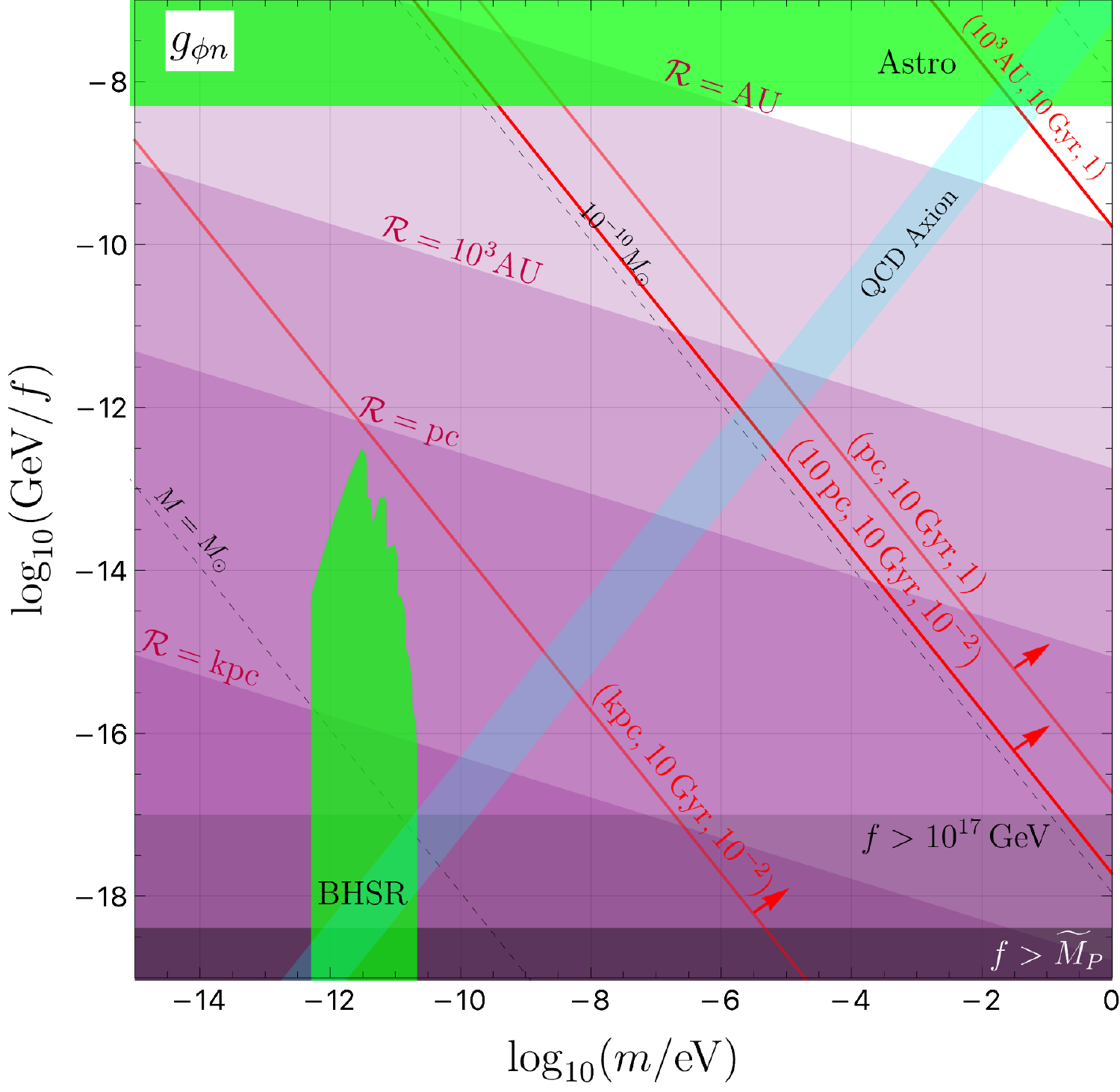}
 \caption{\label{fig:QCDsens_nucleons}
 Regions of parameter space for the axion-nucleon coupling $g_{\phi n}$ where $g_*/g_\textrm{DM}<1$ using Eq.~\eqref{SNR_spatial-derivative_no-wave-spreading} (purple regions), for the different explosion distances $\Rcal=$ 1\,AU, $10^3$\,AU, 1\,pc, 1\,kpc. 
 The red and black lines, as well as the cyan and horizontal gray shaded regions, are the same as in Fig.~2(a) in the Main Text. 
 The green regions denote current astrophysical constraints on $g_{\phi n}$ \cite{Arvanitaki:2014wva,DiLuzio:2021gos}.} 
\end{figure}

In the Main Text, we discussed effects of the axion-photon coupling that involve the time derivative of the axion field. 
Here, we discuss effects of the axion-fermion coupling that involve the spatial gradient of the axion field. 
Let us consider the derivative-type coupling of an axion field to fermions: 
\begin{equation}
\label{Lagrangian_axion-fermion_coupling}
\Lagr_\textrm{fermions} = 
g_{\phi f} \left(\pd_\m \phi\right) \bar{f}\g^\m\g_5 f \, , 
\end{equation}
where $f$ is a fermion field and $\bar{f} = f^\dagger \gamma^0$ is its Dirac adjoint. 
The spatial components in Eq.~(\ref{Lagrangian_axion-fermion_coupling}) give rise to an interaction of the form $H \propto g_{\phi f} \boldsymbol{\nabla} \phi \cdot \boldsymbol{\Sigma}$, where $\boldsymbol{\Sigma}$ is the Dirac spin matrix vector of the fermion. 
Broadband searches for axion DM via the fermion couplings in Eq.~(\ref{Lagrangian_axion-fermion_coupling}) rely on the precession of polarised fermion spins about $\boldsymbol{\nabla} \phi$ \cite{Flambaum:2013-axion,Stadnik:2014axion,Stadnik:2017-axion,Abel:2017-axion,Terrano:2019-axion,Smorra:2019qfx}. 
In this case, the sensitivity to a relativistic axion burst, relative to the standard cold DM search paradigm, is given by: 
\begin{equation}
\label{SNR_spatial-derivative_no-wave-spreading}
\frac{g_*}{g_\textrm{DM}} \sim v_\textrm{DM}  \sqrt{\frac{\rho_\textrm{DM}}{\rho_*}} \frac{ t_
\textrm{int}^{1/4} \textrm{min} \left( \tau_\textrm{DM}^{1/4} , t_\textrm{int}^{1/4} \right)}{ \min \left[ \left( \delta t \right)^{1/4} , t_\textrm{int}^{1/4} \right] \min \left( \tau_*^{1/4} , t_\textrm{int}^{1/4} \right) } \, , \end{equation}
where we have used the inequality $\tau_* < \delta t$, analogously to Eq.~(\ref{SNR_time-derivative_no-wave-spreading}) in the Main Text. 
In this case, the sensitivity to a relativistic axion burst is enhanced by the factor of $v_*/v_\textrm{DM} \sim 10^3$ compared to Eq.~\eqref{SNR_time-derivative_no-wave-spreading}. 

In Fig.~\ref{fig:QCDsens_nucleons}, we show the regions of parameter space where $g_{\ast}/g_{\rm DM} < 1$ for the axion-nucleon coupling $g_{\phi n}$, if a bosenova occurred within a distance $\Rcal$ of the detector during an experimental search duration of $t_{\rm int}=$ 1\,yr. 
At present, the proposals for a broadband search for the axion-nucleon couplings in the relevant mass range are limited \footnote{For example, the CASPEr-Wind experiment \cite{Garcon:2017ixh} is proposed to be operated as a resonant search in the relevant mass range.}.
However, the sensitivity of such a search would be enhanced for the relativistic axion signal, relative to the cold DM case, by the factor of $v_*/v_\textrm{DM} \sim 10^3$, see Eq.~\eqref{SNR_spatial-derivative_no-wave-spreading}. 
We see again that relativistic axion bursts can outperform cold axion DM searches in the large-$f$ region, even for quite distant bursts. 

%%%%%
\section{Non-derivative Axion Couplings}
\label{app:ND}

In the Main Text and in \ref{app:SD}, we have considered the primary couplings searched for in axion experiments, $g_{\phi\gamma}$ and $g_{\phi n}$ in Eqs.~\eqref{Lagrangian_axion-photon_coupling} and \eqref{Lagrangian_axion-fermion_coupling}, respectively, to estimate the sensitivity ratios in Eqs.~\eqref{SNR_time-derivative_no-wave-spreading} and \eqref{SNR_spatial-derivative_no-wave-spreading}. 
For completeness, we also consider non-derivative couplings below. 

The axion's coupling to the gluon field tensor $G$ and its dual $\tilde{G}$,
\begin{equation}
\label{Lagrangian_axion-gluon_coupling}
\mathcal{L} = g_{\phi G} \, \phi \, G_{\mu \nu}^b \tilde{G}^{b \mu \nu} \, , 
\end{equation}
where $b$ is the colour index, gives rise to time-varying electric dipole moments of nucleons \cite{Graham:2011axion} and atoms \cite{Stadnik:2014axion,Ritz:2020axion}, which are non-derivative in nature. 
Additionally, scalar-type non-derivative couplings such as
\begin{equation}
\label{Lagrangian_scalar-fermion_coupling}
\mathcal{L} = \tilde{g}_{\phi f} \, \phi \, \bar{f} f \, , 
\end{equation}
may arise in models of axions with \textit{CP} violation in the quark sector \cite{Moody:1984CP} or in models of scalar-field DM that give rise to apparent variations of the fundamental constants \cite{Stadnik;2015scalar1,Stadnik;2015scalar2}. 
In the case of non-derivative couplings, the sensitivity to a relativistic axion burst, relative to the standard cold DM search paradigm, is given by: 
\begin{equation}
\label{SNR_nonderivative_no-wave-spreading}
\frac{g_*}{g_\textrm{DM}} \sim \frac{\varepsilon}{m} \sqrt{\frac{\rho_\textrm{DM}}{\rho_*}} \frac{ t_
\textrm{int}^{1/4} \textrm{min} \left( \tau_\textrm{DM}^{1/4} , t_\textrm{int}^{1/4} \right)}{ \min \left[ \left( \delta t \right)^{1/4} , t_\textrm{int}^{1/4} \right] \min \left( \tau_*^{1/4} , t_\textrm{int}^{1/4} \right) }  \, . 
\end{equation}

\bibliography{axionbib.bib}

\end{document}